\documentclass[twoside]{article}
\usepackage{l3physics}   
\usepackage{fleqn,tau98}
\usepackage{epsf}%
\usepackage{epsfig}%
\renewcommand{\etal}{{\em{et al.}}}
\newcommand{\mW}{m_W}
\newcommand{\mZ}{m_Z}
\newcommand {\kn}{K^-}

\newcommand {\Be}{{\cal{B}}_{e}}
\newcommand {\Bm}{{\cal{B}}_{\mu}}
\newcommand {\Bp}{{\cal{B}}_{\pi}}
\newcommand {\Bk}{{\cal{B}}_{K}}
\newcommand {\Bl}{{\cal{B}}_{\ell}}
\newcommand {\Bh}{{\cal{B}}_{h}}
\newcommand {\GFMUval}   {(1.16639  \pm 0.00002) \times 10^{-5}   \mathrm{GeV^{-2}}} 
\newcommand {\MTval}     {1776.96^{+0.31}_{-0.27}}   
\newcommand {\TAUTval}   {290.5      \pm 1.0}              
\newcommand {\BRTEval}   {17.81      \pm 0.06}             
\newcommand {\BRTMval}   {17.36      \pm 0.06}             
\newcommand {\BRTPval}   {11.08      \pm 0.13}             
\newcommand {\BRTKval}   {0.695      \pm 0.026}            
\newcommand {\FPIVUDval} {(127.4 \pm 0.1)  \mathrm{MeV}}     %
\newcommand {\FKVUSval}  {(35.18 \pm 0.05) \mathrm{MeV}}     %
\newcommand {\GF}        {G_{\mathrm{F}}}

\newcommand {\MNUTlimNOW}    {36}    
\newcommand {\FMIXlimNOW}    {0.0053} 
\newcommand {\ETAlimNOW}     {-0.030 < {\eta_{\tau\mu}} < 0.052}   
\newcommand {\KAPPAMlimNOW}  {-0.011 < \kappa < 0.017}      
\newcommand {\KAPPAElimNOW}  {|\tilde{\kappa}| < 0.26}     
\newcommand {\MNUTlimTCFa}    {34}    
\newcommand {\FMIXlimTCFa}    {0.0039} 
\newcommand {\ETAlimTCFa}     {-0.030 < {\eta_{\tau\mu}} < 0.029}     
\newcommand {\KAPPAMlimTCFa}  {-0.011 < \kappa < 0.009}      
\newcommand {\KAPPAElimTCFa}  {|\tilde{\kappa}| < 0.20}     
\newcommand {\MNUTlimTCFb}    {28}    
\newcommand {\FMIXlimTCFb}    {0.0024} 
\newcommand {\ETAlimTCFb}     {-0.017 < {\eta_{\tau\mu}} < 0.016}  
\newcommand {\KAPPAMlimTCFb}  {-0.006 < \kappa < 0.005}      
\newcommand {\KAPPAElimTCFb}  {|\tilde{\kappa}| < 0.15}     
\title{Sensitivities of one-prong tau branching fractions to   
      tau neutrino mass, mixing, and 
      anomalous charged current couplings\\
       {\normalsize\em{Invited talk at the tau-charm Workshop, 
                       6-9 March 1999, SLAC, USA}}}

\author{Maria Teresa Dova\address{Universidad Nacional de La Plata, La Plata, Argentina},
        John Swain$^{\mathrm{b}}$ and 
        Lucas Taylor\address{Department of Physics,
                             Northeastern University, Boston, MA02115, USA}%
}%

\begin{document}

\begin{abstract}

We analyse the sensitivities of exclusive one-prong tau branching fractions to: 
the tau neutrino mass;
its mixing with a fourth generation neutrino;
the weak charged current magnetic and electric dipole moments of the tau;
and the Michel parameter $\eta$.
Quantitative constraints are derived from current experimental data 
and the future constraints derivable from tau-charm factory measurements
are estimated.
The anomalous coupling constraints are used to constrain the tau compositeness scale 
and the allowed parameter space for Higgs doublet models.

\end{abstract}

\maketitle
\section{INTRODUCTION}

We analyse the sensitivity to new physics of the 
$\tau$ partial 
widths for the following decays%
\footnote{Throughout this paper the charge-conjugate decays are also implied.
          We denote the branching ratios for these processes as
          $\Be, \Bm, \Bp, \Bk$ respectively;
          $\Bl$ denotes either $\Be$ or $\Bm$ while $\Bh$ denotes either $\Bp$ or $\Bk$.}
:
$\tau^-\rightarrow{e}^-\bar{\nu}_{{e}}\nu_\tau$,
$\tau^-\rightarrow\mu^-\bar{\nu}_\mu\nu_\tau$,
$\tau^-\rightarrow\pi^-\nu_\tau$, and
$\tau^-\rightarrow{K}^-\nu_\tau$.
We determine constraints on the mass $m_{\nu_3}$ of the third generation 
neutrino $\nu_3$, its mixing with a fourth generation 
neutrino $\nu_4$ of mass $ > M_Z/2$,
anomalous weak charged current magnetic and 
electric dipole couplings~\cite{RIZZO97A,CHIZHOV96A},
and the Michel parameter $\eta$~\cite{MICHEL}. 
In each case, we present quantitative results using current experimental 
data (which update our previous analyses~\cite{MNUTAU,MASS_MIX_UPDATE,ANOMALOUS_COUPLINGS}) 
and estimate the future constraints which would be achievable 
using the expected precision of measurements at a tau-charm factory.
The results for the $\eta$ parameter are used to constrain
extensions of the Standard Model which contain more than one 
Higgs doublet and hence charged Higgs bosons.

%
\section{THEORETICAL PREDICTIONS}
%
\subsection{Tau neutrino mass and mixing}

The theoretical predictions for the branching fractions $\Bl$ 
allowing for the $\nu_\tau$ mass and mixing with a fourth lepton
generation are given by~\cite{MNUTAU}:
\begin{eqnarray}
\Bl^{\mathrm{th.}}\!&\!=\!&\!\frac {\GF^2 m_\tau^5\tau_\tau}{192\pi^3}\left( 1 -  8x - 12 x^2{\mathrm{ln}}x + 8 x^3 - x^4\right)                 \nonumber \\                                      
              &\!\times\!& \!\!\!\left[\left(
                                          1 - \frac{\alpha(m_\tau)}{2\pi} 
                                          \left( 
                                                 \pi^2 - \frac{25}{4} 
                                          \right) 
                            \right)  
                           \left(1 + \frac{3}{5} \frac{m_\tau^2}{\mW^2} 
                           \right) \right]                                             \nonumber  \\
             &\!\times\!&\!\!\!\left[ 1 - \sin^2\theta \right] \left[ 1 - 8y(1-x)^3+\cdots\right]
\label{equ:bleptm}
\end{eqnarray}
where $x=m_\ell^2/m_\tau^2$, $y=m_{\nu_3}^2 / m_\tau^2$,
$\GF    = \GFMUval$ is the Fermi constant~\cite{PDG96SHORT}, and 
$\tau_\tau$ is the tau lifetime.
The tau mass, $m_\tau$, is taken only from production measurements 
at tau-pair threshold since values derived from kinematic reconstruction of 
tau decays depend on tau neutrino mass.
The first term in square brackets allows for radiative 
corrections\cite{BERMAN58A,KINOSHITA59A,SIRLIN78A,MARCIANO88A}, 
where $\alpha(m_\tau)\simeq 1/133.3$ is the QED coupling constant~\cite{MARCIANO88A} 
and $\mW = 80.41 \pm 0.10$\,GeV is the $W$ mass~\cite{PDG98}.

The tau neutrino weak eigenstate is given by the
superposition of two mass eigenstates 
$|\nu_\tau\rangle = \cos\theta |\nu_3\rangle + \sin\theta |\nu_4\rangle$, 
such that the mixing is parametrised by the Cabibbo-like mixing angle $\theta$.
The second term in square brackets describes mixing with a fourth generation neutrino
which, being kinematically forbidden, causes a suppression of the decay rate.
The third term in brackets  parametrises the suppression 
due to a non-zero mass of $\nu_3$, where the ellipsis denotes negligible higher order 
terms~\cite{MNUTAU}.                              

The branching fractions for the decays $\tau^-\rightarrow{h}^-\nu_\tau$, 
with ${h}=\pi/{K}$, are given by~\cite{MNUTAU}
\begin{eqnarray}  
\Bh^{\mathrm{th.}}\!\!\!\!&\!=\!&\!\!\!\!\left(\frac {\GF^2 m_\tau^3 } {16\pi}\right)\tau_\tau f_{{h}}^2 |V_{\alpha\beta}|^2  
                       \left(1 - x\right)^2                                                          \nonumber \\         
                 &\!\times\!&\!\!\!\!\!\left( 
                                     1 + \frac{2\alpha}{\pi} {\mathrm{ln}} 
                                     \left( \frac {\mZ} {m_\tau} \right)+\cdots 
                       \right) \left[ 1 - \sin^2\theta  \right]                                              \nonumber  \\                                                                                                                                             
             &\!\times\!&\!\!\!\!\!\left[1\!-\!y\left(\frac{2\!+\!x\!-\!y}{1-x}\right)\left(1\!-\!\frac{y(2\!+\!2x\!-\!y)}{(1-x)^2}\right)^{\frac{1}{2}}\right]               
         \label{equ:bhad}
\end{eqnarray}
where 
$x=m_{{h}}^2 / m_\tau^2$,
$m_{{h}}$ is the hadron mass, 
$f_{{h}}$ are the hadronic form factors, 
and $V_{\alpha\beta}$ are the CKM matrix elements, 
$V_{{ud}}$ and $V_{{us}}$,
for $\pi^-$ and $\kn$ respectively.
The quantities 
$f_\pi |V_{{ud}}|   = \FPIVUDval$ and 
$f_{{K}} |V_{{us}}| = \FKVUSval$ are obtained from 
analyses of $\pi^-\rightarrow\mu^-\bar{\nu}_\mu$ and
$\kn\rightarrow\mu^-\bar{\nu}_\mu$ decays~\cite[and references therein]{MARCIANO92A}.
The ellipsis represents terms, estimated to be ${\cal{O}}(\pm 0.01)$\cite{MARCIANO92A},        
which are neither explicitly treated nor implicitly absorbed into $\GF$,
$f_\pi |V_{{ud}}|$, or $f_{{K}} |V_{{us}}|$.
The first term in square brackets describes mixing with a fourth generation neutrino
while the second parametrises the effects of 
a non-zero $m_{\nu_3}$.

The fourth generation neutrino mixing affects all the tau branching
fractions with a common factor whereas a non-zero tau neutrino 
mass affects all channels with different kinematic factors.
Therefore, given sufficient experimental precision, these two effects 
could in principle be separated.

Analyses which determine the tau mass from a kinematic reconstruction of the 
tau decay products are also sensitive to tau neutrino mass.
For example, from an analysis of 
$\tau^+\tau^-$ 
$\rightarrow$ 
$(\pi^+n\pi^0\bar{\nu}_\tau)$
$(\pi^-m\pi^0\nu_\tau)$ 
events (with $n\leq2, m\leq2, 1\leq n+m\leq3$),
CLEO determined the $\tau$ mass to be 
$m_\tau = (1777.8 \pm 0.7 \pm 1.7) + [m_{\nu_3}({\mathrm{MeV}})]^2/1400$ MeV\cite{CLEOWEINSTEIN}.
Such measurements may be used to further constrain $m_{\nu_3}$.

\subsection{Anomalous couplings}
The theoretical predictions for the branching fractions $\Bl$ for the 
decay $\tau^-\rightarrow\ell^-\bar{\nu}_{\ell}\nu_\tau (X_{\mathrm{EM}})$, with
$\ell^-=\mathrm{e}^-, \mu^-$ and $X_{\mathrm{EM}} = \gamma,~\gamma\gamma,~e^+e^-,\ldots$, 
are given by:
\begin{eqnarray}
  \Bl^{\mathrm{th.}}\!\!&\!\!=\!\!&\!\frac {\GF^2 m_\tau^5\tau_\tau}{192\pi^3}
                             \left( 1 -  8x - 12 x^2{\mathrm{ln}}x + 8 x^3 - x^4\right) \nonumber \\                      
              &\!\times\!& \!\!\left(
                                          1 - \frac{\alpha(m_\tau)}{2\pi} 
                                          \left( 
                                                 \pi^2 - \frac{25}{4} 
                                          \right) 
                            \right)  
                           \left(1 + \frac{3}{5} \frac{m_\tau^2}{\mW^2} 
                           \right)                                               \nonumber  \\ 
              &\!\times\!& \left[ 1 + \Delta_\ell \right].
\label{equ:blept}
\end{eqnarray}
The term in square brackets describes the effects of new physics 
where the various $\Delta_\ell$ we consider are defined below.

The effects of anomalous weak charged current dipole moment 
couplings at the $\tau\nu_\tau W$ vertex are 
described by the effective Lagrangian
\begin{eqnarray}
{\cal{L}}\!&\!=\!&\!\frac{g}{\sqrt{2}}\bar{\tau} 
                            \left[ 
                            \gamma_\mu +
                            \frac{i\sigma_{\mu\nu}q^\nu}{2m_\tau}
                            (\kappa_\tau-i\tilde{\kappa}\gamma_5)
                            \right]
                      P_L \nu_\tau W^\mu \nonumber \\
                      & & + ({\mathrm{Hermitian\ conjugate}}),
\end{eqnarray}
where $P_L$ is the left-handed projection 
operator and the parameters $\kappa$ and 
${\tilde{\kappa}}$ are the (CP-conserving) magnetic and (CP-violating) electric dipole form 
factors respectively~\cite{RIZZO97A}.
They are the charged current analogues of the weak neutral current dipole 
moments, measured using $Z\rightarrow\tau^+\tau^-$ events~\cite{PICH97A}, 
and the electromagnetic dipole moments~\cite{BIEBEL96A,TTGNUCPHYSB} 
recently measured by L3 and OPAL using 
$Z\rightarrow\tau^+\tau^-\gamma$ events~\cite{OPALTTG,L3TTG,TAYLOR_TAU98}.
In conjunction with Eq.~\ref{equ:blept}, the effects of non-zero 
values of $\kappa$ and ${\tilde{\kappa}}$ on the tau leptonic 
branching fractions may be described by~\cite{RIZZO97A}
\begin{eqnarray}
  \Delta_\ell^{\kappa}         & = &  {\kappa}/{2} + {\kappa^2}/{10};\label{equ:kappa}     \\         
  \Delta_\ell^{\tilde{\kappa}} & = &  {\tilde{\kappa}^2}/{10}.       \label{equ:kappatilde}        
 \label{equ:deltalk}
\end{eqnarray}
The dependence of the tau leptonic branching ratios on $\eta$ is given,
in conjunction with Eq.~\ref{equ:blept}, by~\cite{STAHL94A}
\begin{eqnarray}
  \Delta_\ell^{\eta} & = &  4{\eta_{\tau\ell}} {\sqrt{x}},                  
\label{equ:deltaln}
\end{eqnarray}
where the subscripts on $\eta$ denote the initial and final 
state charged leptons.
%
Both leptonic tau decay modes probe the charged current couplings of 
the transverse $W$, and are sensitive to $\kappa$ and ${\tilde{\kappa}}$.
In contrast, only the $\tau^-\rightarrow\mu^-\bar{\nu}_\mu\nu_\tau$ channel 
is sensitive to $\eta$, due to a relative suppression factor of $m_e/m_\mu$ for 
the $\tau^-\rightarrow\mathrm{e}^-\bar{\nu}_{\mathrm{e}}\nu_\tau$ channel.
Semi-leptonic tau branching fractions are not considered since 
they are insensitive to $\kappa$,  ${\tilde{\kappa}}$, and $\eta$.

\section{RESULTS}

Three sets of fits are performed, as follows.
\begin{itemize}
\item {\bf{Case 1}} \\
       We use current world averages of the experimental measurements.
\item {\bf{Case 2}}\\
      We use estimated errors on measurements which would be possible 
      with a tau-charm factory assuming that there
      is no improvement in the tau lifetime compared to current measurements.
\item {\bf{Case 3}} \\
      This is identical to Case 2 except that, in order to assess the 
      limiting factors of our method, we assume somewhat arbitrarily  
      that CLEO and the b-factories succeed in reducing the tau-lifetime 
      error by a factor of two.
\end{itemize}
For Cases 2 and 3 the central values are clearly unknown,
therefore in making our predictions we adjust the branching 
fractions to their standard model values, such that our
predictions is not arbitrarily biased by the current 
experimental central values.
The input parameters for the three cases are summarised in
Tab.~\ref{tab:inputpars}.

%
\renewcommand{\arraystretch}{1.3}
\begin{table}[!htbp]
\caption{Input parameters used in the determinations of 
         $m_{\nu_3}$,
         $\sin^2\theta$,          
         $\kappa$,           
         $\tilde\kappa$, and     
         ${\eta_{\tau\mu}}$.}
\label{tab:inputpars}   
{
\small
\setlength{\tabcolsep}{0.2em}
\begin{tabular}{|c|cc|cc|cc|}  \hline 	                                 
%
%
                  & 
\multicolumn{2}{c|}{Value} &
\multicolumn{4}{c|}{Future Error}        \\ \cline{4-7}  			
                  & 
\multicolumn{2}{c|}{Case 1} &
\multicolumn{2}{c|}{Case 2} &
\multicolumn{2}{c|}{Case 3}        \\ \hline 			
$m_\tau$ (MeV)    & $\MTval$   &   \cite{MTAUBESNEW}        & 0.1      & \cite{TAUCHARM93_PICH}   & 0.1      & \cite{TAUCHARM93_PICH}      \\
$\tau_\tau$ (fs)  & $\TAUTval$ &   \cite{TAU98_WASSERBAECH} & $1.0$    & \cite{TAU98_WASSERBAECH} & $0.5$    &                             \\
$\Be$ (\%)        & $\BRTEval$ &   \cite{TAU98_STUGU}       & 0.018    & \cite{TAUCHARM93_PICH}   & 0.018    & \cite{TAUCHARM93_PICH}      \\
$\Bm$ (\%)        & $\BRTMval$ &   \cite{TAU98_STUGU}       & 0.017    & \cite{TAUCHARM93_PICH}   & 0.017    & \cite{TAUCHARM93_PICH}      \\
$\Bp$ (\%)        & $\BRTPval$ &   \cite{PDG98}             & 0.011    & \cite{TAUCHARM93_PICH}   & 0.011    & \cite{TAUCHARM93_PICH}      \\
$\Bk$ (\%)        & $\BRTKval$ &   \cite{TAU98_HELTSLEY}    & 0.003    & \cite{TAUCHARM93_PICH}   & 0.003    & \cite{TAUCHARM93_PICH}      \\ \hline
\end{tabular}
}
\end{table}

We derive constraints on $m_{\nu_\tau}$ and $\sin^2\theta$
from combined likelihood fits to the four tau decay channels,
using equations \ref{equ:bleptm} and \ref{equ:bhad}.
The likelihood for the CLEO and BES measurements of $m_\tau$ 
to agree, as a function of 
$m_{\nu_3}$, is included in the global likelihood. 
We derive constraints on 
$\kappa$,           
$\tilde\kappa$, and     
${\eta_{\tau\mu}}$
using the two leptonic tau decay channels and 
Eq.~\ref{equ:blept}.
Each of the five parameters is analysed separately, conservatively 
assuming in each case that the other four parameters are zero.

In the fit, the uncertainties on all the quantities in 
Eqs.~\ref{equ:bleptm},~\ref{equ:bhad}, and 
~\ref{equ:blept} are taken into account.
The likelihood is constructed numerically following the 
procedure of Ref.~\cite{NIM_LIKELIHOOD_PAPER} by randomly 
sampling all the quantities used according to their errors.

Tab.~\ref{tab:results} summarises the results obtained.
%
\renewcommand{\arraystretch}{1.3}
\begin{table*}[!htbp]
\caption{Constraints on 
         $m_{\nu_3}$,
         $\sin^2\theta$,          
         $\kappa$,           
         $\tilde\kappa$, and     
         ${\eta_{\tau\mu}}$ at the 95\% confidence level.}
\label{tab:results}   
{
\setlength{\tabcolsep}{1.0em}
\begin{tabular}{|c|c|c|}   \hline 	                                
 Case 1                                & Case 2                         & Case 3                          \\ \hline
 $m_{\nu_3}       <  \MNUTlimNOW$ MeV  & $m_{\nu_3}       <  \MNUTlimTCFa$ MeV  & $m_{\nu_3}       <  \MNUTlimTCFb$ MeV   \\
 $\sin^2\theta    <  \FMIXlimNOW$      & $\sin^2\theta    <  \FMIXlimTCFa$      & $\sin^2\theta    <  \FMIXlimTCFb$       \\
 ${\KAPPAMlimNOW}$                     & ${\KAPPAMlimTCFa}$                     & ${\KAPPAMlimTCFb}$                      \\
 ${\KAPPAElimNOW}$                     & ${\KAPPAElimTCFa}$                     & ${\KAPPAElimTCFb}$                      \\
 ${\ETAlimNOW}$                        & ${\ETAlimTCFa}$                        & ${\ETAlimTCFb}$                         \\ \hline
\end{tabular}
}
\end{table*}
%
For Cases 2 and 3 the limiting error is that on the tau lifetime;
arbitrarily setting all other errors 
to zero yields negligible improvement in the fit results.

\section{DISCUSSION}
%
\subsection{Tau neutrino mass}

The limit on $m_{\nu_3}$ can be reasonably interpreted as a 
limit on $m_{\nu_\tau}$, since $\sin^2\theta$ is small as well as the mixing of 
$m_{\nu_3}$ with lighter neutrinos~\cite{PDG98}.
The best direct experimental constraint on the tau neutrino mass 
is $m_{\nu_\tau} < 18.2$\,MeV at the 95\% confidence level\cite{ALEPH_NUTAU} 
which was obtained using many-body hadronic decays of the $\tau$.
While our constraint is less stringent, it is statistically independent.
Moreover, it is insensitive to fortuitous or pathological events 
close to the kinematic limits, the absolute energy scale of the detectors,
and the details of the resonant structure of multi-hadron $\tau$ decays~\cite{TAU98_MCNULTY}.

Although the constraint on $m_{\nu_\tau}$ which we estimate does improve
with the tau-charm input, this method would not be competitive with 
direct reconstruction analyses which are predicted to be sensitive at the
$O(2\,{\mathrm{MeV}})$ level~\cite{TAUCHARM93_CADENAS}.

\subsection{Fourth generation mixing}

Our upper limit on $\sin^2\theta$ is already the most stringent 
experimental constraint on mixing of the third and fourth 
neutrino generations.
This constraint will improve by a factor of up to two
using future tau-charm factory data, depending on the 
improvement in the error on $\tau_\tau$.
We anticipate that this technique will continue to provide the most 
stringent constraints in the foreseeable future.
  
\subsection{Anomalous couplings and tau compositeness}
Our results for $\kappa$ and $\tilde\kappa$ are currently the most precise.
The less stringent constraint on $\tilde\kappa$ compared to that
on $\kappa$ is due to the lack of linear terms
in Eq.~\ref{equ:kappatilde}.

Derivative couplings necessarily involve the introduction of a length
or mass scale. 
Anomalous magnetic moments due to compositeness are expected to be
of order $m_\tau/\Lambda$ where $\Lambda$ is the compositeness 
scale~\cite{BRODSKY80A}.
We can then interpret the 95\% confidence level on $\kappa$, 
the quantity for which we have a more stringent bound, as a statement
that the $\tau$ appears to be a point-like Dirac particle up to an energy
scale of $\Lambda \approx m_\tau/0.017 = 105$\,GeV. 
These results are comparable to those obtained from anomalous weak 
neutral current couplings~\cite{PICH97A} and more stringent than those 
obtained for anomalous electromagnetic couplings~\cite{OPALTTG,L3TTG,TAYLOR_TAU98}.
While the decay $W\rightarrow \tau\nu$ which is measured at LEP\,II 
is also sensitive to charged current dipole terms, given that the 
energy scale is $\mW$, the interpretation in terms of the static 
properties $\kappa$ and $\tilde\kappa$ is less clear.

The results for $\kappa$ and $\tilde\kappa$ will improve by using tau-charm 
data, and will probe the point-like nature of the tau up to a scale of 
$\Lambda = O(180\,{\mathrm{GeV}})$ (assuming no improvement in $\tau_\tau$) or 
$\Lambda = O(300\,{\mathrm{GeV}})$ (assuming a factor of two improvement in 
the error on $\tau_\tau$).   

\subsection{${\eta_{\tau\mu}}$ and extended Higgs sector models}
Our value for ${\eta_{\tau\mu}}$ is currently the most precise.
The uncertainty is significantly smaller than determinations using 
the shape of momentum spectra of muons from $\tau$ decays, 
$({\eta_{\tau\mu}} = -0.04 \pm 0.20)$~\cite{PICH97A}.

\begin{figure}[htb]
\epsfig{file=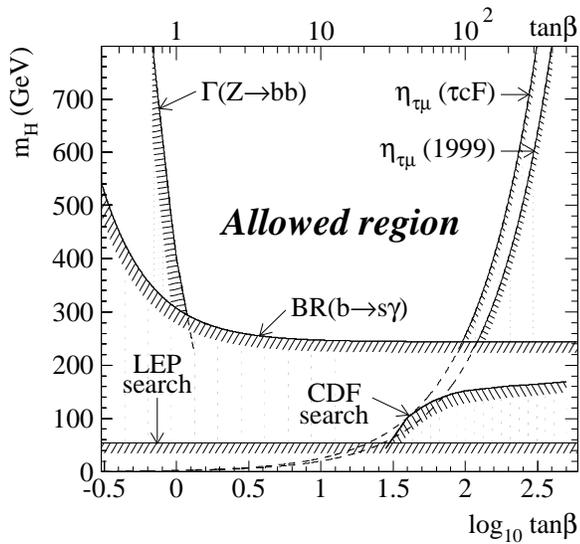,width=0.48\textwidth,clip=}\\[-4mm]
\caption{Constraints on $m_H$ as a function of $\tan\beta$ at the 95\% C.L.,
         from this analysis of $\eta_{\tau\mu}$ and the other analyses 
         described in the text.}\label{fig:chhiggs}
\end{figure}

Many extensions of the Standard Model, such as supersymmetry (SUSY),
involve an extended Higgs sector with more than one Higgs doublet. 
Such models contain charged Higgs bosons which contribute 
to the weak charged current with couplings which depend on the fermion masses.
Of all the Michel parameters, ${\eta_{\tau\mu}}$ is especially
sensitive to the exchange of a charged Higgs.
Following Stahl~\cite{STAHL94A}, ${\eta_{\tau\mu}}$ can be written as 
\begin{equation}
{\eta_{\tau\mu}} = -\left( \frac{m_\tau m_\mu}{2} \right)
        \left( \frac{\tan\beta}{m_H} \right)^2
\end{equation}
where $\tan\beta$ is the ratio of vacuum expectation values of the
two Higgs fields, and $m_H$ is the mass of the charged Higgs.
This expression applies to type II extended Higgs sector models 
in which the up-type quarks get their masses from one doublet and 
the down-type quarks get their masses from the other. 
From current data we determine the one-sided constraint ${\eta_{\tau\mu}} > -0.0232$ 
at the 95\% C.L. which rules out the region
    $m_H < (2.01 \tan\beta) \,{\mathrm{GeV}}$ at the 95\% C.L.
as shown in Fig.~\ref{fig:chhiggs}.

An almost identical constraint on the high $\tan\beta$ region of 
type II models may be obtained from the process 
$B\rightarrow\tau\nu$~\cite{HOU93A}.
The most stringent constraint, from the L3 experiment,
rules out the region
    $m_H < (2.09 \tan\beta)\,{\mathrm{GeV}}$ at the 95\% C.L.~\cite{L3B2TAU}.
Within the specific framework of the minimal supersymmetric standard 
model, the process $B\rightarrow\tau\nu X$ rules out the region 
$m_H < (2.33 \tan\beta)\,{\mathrm{GeV}}$ at the 
95\% C.L.~\cite{COARASA97A}.
This limit, however, depends on the value of the Higgsino mixing 
parameter $\mu$ and can be evaded completely for $\mu>0$. 
The non-observation of proton decay also tends to rule out 
the large $\tan\beta$ region but these constraints are 
particularly model-dependent.
The very low $\tan\beta$ region is ruled out by measurements of 
the partial width $\Gamma (Z\rightarrow b\bar{b})$. 
For type II models the approximate region excluded is 
$\tan\beta < 0.7$ at the $2.5\sigma$ C.L. for 
any value of $M_H$~\cite{GRANT95A}.  
Complementary bounds for the full $\tan\beta$ region are derived 
from the CLEO measurement of 
$BR(b\rightarrow s\gamma) = (2.32\pm0.57\pm 0.35)\times 10^{-4}$  
which rules out, for type II models, the region 
$M_H < 244 + 63/{(\tan\beta)}^{1.3}$~\cite{CLEO_BTOSGAMMA}.
This constraint can, however, be circumvented in SUSY models
where other particles in the loops can cancel out the effect
of the charged Higgs.
Direct searches at LEP II exclude the region $m_H < 54.5$\,GeV for 
all values of $\tan\beta$~\cite{DELPHIMH}.
The CDF search for charged Higgs bosons in the process
$t\rightarrow b H^+$ rules out the region of low $m_H$ and 
high $\tan\beta$~\cite{ABE97A}.
The 95\% C.L. constraints in the $m_H$ {\em{vs.}} $\tan\beta$ plane, 
from this and other analyses, are shown in Fig.~\ref{fig:chhiggs}.

%

We anticipate that the constraints from $Z\rightarrow b\bar{b}$ 
and $b\rightarrow s \gamma$ will improve somewhat with new 
measurements from LEP, CLEO, and the b-factories and 
from refinements in the theoretical treatment~\cite{CIUCHINI97A}. 
CLEO and the b-factories may also improve the measurements of 
$B\rightarrow\tau\nu(X)$ which rule out a similarly-shaped region 
of the $m_H-\tan\beta$ plane as that of this analysis.

Some caution is advised in the interpretation of the large $\tan\beta$ 
regime which becomes non-perturbative for $\tan\beta > O(70)$. 
Future improved measurements of the tau branching fractions and 
lifetime will, however, extend the constraints on $\tan\beta$ towards 
lower values, where perturbative calculations are more applicable.

In particular, for the tau-charm factory we estimate 
the one-sided constraint ${\eta_{\tau\mu}} > -0.014$ 
at the 95\% C.L.
This rules out the region
    $m_H < (2.55 \tan\beta) \,{\mathrm{GeV}}$ at the 95\% C.L.,
as shown in Fig.~\ref{fig:chhiggs},
and corresponds to $\sim$25\% reduction in the maximum allowed 
value of $\tan\beta$ for a given value of $m_H$, compared 
to current constraints.

\section{SUMMARY}
From an analysis of tau leptonic and semileptonic
branching fractions we determine constraints on 
$m_{\nu_\tau}$,
$\sin^2\theta$,     
$\kappa$,           
$\tilde\kappa$, and 
${\eta_{\tau\mu}}$ 
using current experimental data.
We then assess the future sensitivity to these parameters 
using predictions for the uncertainties on experimental 
quantities measured at a tau-charm factory.
We find that in each case the future sensitivity is completely  
limited by the uncertainty on the tau lifetime.

The constraint on $m_{\nu_\tau}$ using current data is 
complementary to, but less stringent than, that already 
obtained from multi-hadronic tau decays.
Our technique will benefit slightly from improved tau-charm 
factory data but will be considerably less competitive than 
other techniques available at such a facility.

Using current experimental data we find that our technique 
yields the most stringent constraints to-date on 
$\sin^2\theta$,     
$\kappa$,           
$\tilde\kappa$, and 
${\eta_{\tau\mu}}$.
All these constraints are expected to improve by a factor of 
approximately two using future data from a tau-charm factory
and, and in the absence of novel competing techniques, will 
continue to yield the most precise determinations of these 
quantities.

The result for $\kappa$ indicates that the tau is point-like up
to an energy scale of approximately 105\,GeV (today) and 
$O(300\,{\mathrm{GeV}})$ (using tau-charm data and assuming a 
factor of two improvement in the tau-lifetime error).

The result for ${\eta_{\tau\mu}}$ constrains the charged Higgs 
of type II two-Higgs doublet models such that we can exclude, 
at the 95\% C.L., the region
$m_H < (2.01 \tan\beta) \,{\mathrm{GeV}}$ (today) 
and
$m_H < (2.55 \tan\beta) \,{\mathrm{GeV}}$ (using tau-charm data and assuming a 
factor of two improvement in the tau-lifetime error). 

\section*{Acknowledgements}
J.S. would like to thank the organisers and participants for 
a stimulating and productive workshop.
We would like to thank CONICET, Argentina (M.T.D.) and 
the NSF, USA (J.S. and L.T.) for financial support.
   

\begin{thebibliography}{99}

\bibitem{RIZZO97A}
T.G. Rizzo.
\newblock {\em Phys. Rev.}, D56:3074, 1997.

\bibitem{CHIZHOV96A}
{M.V. Chizhov}.
\newblock {{\bf{hep-ph/9612399}} (unpublished)}.

\bibitem{MICHEL}
L.~Michel.
\newblock {\em Proc. Phys. Soc.}, A63:514, 1950.

\bibitem{MNUTAU}
J.~Swain and L.~Taylor.
\newblock {\em Phys. Rev.}, D 55:1\,R, 1997.

\bibitem{MASS_MIX_UPDATE}
{J. Swain, and L. Taylor}.
\newblock {\bf{hep-ph/9712383}}.

\bibitem{ANOMALOUS_COUPLINGS}
{M.T. Dova, J. Swain and L. Taylor}.
\newblock {\em Phys. Rev.}, D 58:015005, 1998.

\bibitem{PDG96SHORT}
{R. M. Barnett {\em{et al.}}}
\newblock {\em Phys. Rev.}, D54:1, 1996.

\bibitem{BERMAN58A}
S.~M. Berman.
\newblock {\em Phys. Rev.}, 112:267, 1958.

\bibitem{KINOSHITA59A}
T.~Kinoshita and A.~Sirlin.
\newblock {\em Phys. Rev.}, 113:1652, 1959.

\bibitem{SIRLIN78A}
A.~Sirlin.
\newblock {\em Rev. Mod. Phys.}, 50:573, 1978.

\bibitem{MARCIANO88A}
W.~J. Marciano and A.~Sirlin.
\newblock {\em Phys. Rev. Lett.}, 61:1815, 1988.

\bibitem{PDG98}
{Particle Data Group, C. Caso {\em{et al.}}}
\newblock {\em European Phys. Journ.}, C3:1, 1998.

\bibitem{MARCIANO92A}
W.~J. Marciano.
\newblock {\em Phys. Rev.}, D45:R\,721, 1992.

\bibitem{CLEOWEINSTEIN}
R.~Balest et~al.
\newblock {\em Phys. Rev.}, D47:R3671, 1993.
\newblock The neutrino mass dependence quoted in this paper is in error due to
  a typographical oversight (A. Weinstein, private communication). This does
  not, however, affect any other numbers in the paper.

\bibitem{PICH97A}
{A. Pich}.
\newblock {{\bf{hep-ph/9704453}} (1997). To appear in ``Heavy Flavours II'',
  World Scientific, Singapore, Eds. A.J. Buras and M. Lindner}.

\bibitem{BIEBEL96A}
J.~Biebel and T.~Riemann.
\newblock {\em Z. Phys.}, C76:53, 1997.

\bibitem{TTGNUCPHYSB}
{S.S. Gau, T. Paul, J. Swain, and L. Taylor}.
\newblock {\em Nucl. Phys.}, B 523:439, 1998.

\bibitem{OPALTTG}
{OPAL Collab., K. Ackerstaff \etal}.
\newblock {\em Phys. Lett.}, B431:188, 1998.

\bibitem{L3TTG}
{L3 Collab., M. Acciarri {\em{et al.}}}
\newblock {\em Phys. Lett.}, B434:169, 1998.

\bibitem{TAYLOR_TAU98}
{L. Taylor}.
\newblock {Anomalous magnetic and electric dipole moments of the tau}.
\newblock In {A. Pich and A. Ruiz}, editor, {\em Proceedings of the TAU\,98
  Workshop}, Santander, Spain, 14-17 September 1998. {to appear in: Nucl. Phys.
  B (Proc. Suppl.)}.
\newblock {\bf{hep-ph/9810463}}.

\bibitem{STAHL94A}
A.~Stahl.
\newblock {\em Phys. Lett.}, B324:121, 1994.

\bibitem{MTAUBESNEW}
J.Z. Bai.
\newblock {\em Phys. Rev.}, D53:20, 1996.

\bibitem{TAUCHARM93_PICH}
{A. Pich}.
\newblock {Tau physics prospects at the Tau-Charm Factory and at other
  machines}.
\newblock In J.~Kirkby and R.~Kirkby, editors, {\em Third Workshop on the
  Tau-Charm Factory}, Marbella, Spain, 1-6 June 1993. {Editions
  Froni{\`{e}}res, Gif-sur-Yvette Cedex, France}.
\newblock {page 57}.

\bibitem{TAU98_WASSERBAECH}
{S. Wasserbaech}.
\newblock {Review of tau lifetime measurements}.
\newblock In A.~Pich and A.~Ruiz, editors, {\em Proceedings of the TAU\,98
  Workshop}, Santander, Spain, 14-17 September 1998. {to appear in: Nucl. Phys.
  B (Proc. Suppl.)}.

\bibitem{TAU98_STUGU}
{B. Stugu}.
\newblock {Summary on Tau Leptonic Branching Ratios and Universality}.
\newblock In A.~Pich and A.~Ruiz, editors, {\em Proceedings of the TAU\,98
  Workshop}, Santander, Spain, 14-17 September 1998. {to appear in: Nucl. Phys.
  B (Proc. Suppl.)}.

\bibitem{TAU98_HELTSLEY}
{B. Heltsley}.
\newblock {Experimental Summary on Hadronic Decays}.
\newblock In A.~Pich and A.~Ruiz, editors, {\em Proceedings of the TAU\,98
  Workshop}, Santander, Spain, 14-17 September 1998. {to appear in: Nucl. Phys.
  B (Proc. Suppl.)}.

\bibitem{NIM_LIKELIHOOD_PAPER}
{J. Swain, and L. Taylor}.
\newblock {\em Nucl. Instrum. \& Methods.}, A 411:153, 1998.

\bibitem{ALEPH_NUTAU}
{ALEPH Collab., R. Barate \etal}.
\newblock {\em Eur. Phys. J.}, C2:395, 1998.

\bibitem{TAU98_MCNULTY}
{R. McNulty}.
\newblock {Review of tau neutrino masses}.
\newblock In A.~Pich and A.~Ruiz, editors, {\em Proceedings of the TAU\,98
  Workshop}, Santander, Spain, 14-17 September 1998. {to appear in: Nucl. Phys.
  B (Proc. Suppl.)}.

\bibitem{TAUCHARM93_CADENAS}
{J. G{\'{o}}mez-Cadenas}.
\newblock {Sensitivity of Future ${\mathrm{e^+e^-}}$ colliders to a massive
  $\tau$ Neutrino}.
\newblock In J.~Kirkby and R.~Kirkby, editors, {\em Third Workshop on the
  Tau-Charm Factory}, Marbella, Spain, 1-6 June 1993. {Editions
  Froni{\`{e}}res, Gif-sur-Yvette Cedex, France}.

\bibitem{BRODSKY80A}
{S.J. Brodsky and S.D. Drell}.
\newblock {\em Phys. Rev.}, D22:2236, 1980.

\bibitem{HOU93A}
{W.-S. Hou}.
\newblock {\em Phys. Rev.}, D48:2342, 1993.

\bibitem{L3B2TAU}
{L3 Collab., M. Acciarri \etal}.
\newblock {\em Phys. Lett.}, B396:327, 1997.
\newblock {A. Kunin (private communication)}.

\bibitem{COARASA97A}
{J. A. Coarasa, R. A. Jim\'enez, and J. Sol\'a}.
\newblock {\em Phys. Lett.}, B406:337, 1997.

\bibitem{GRANT95A}
A.~K. Grant.
\newblock {\em Phys. Rev.}, D51:207, 1995.

\bibitem{CLEO_BTOSGAMMA}
{CLEO Collab., M.S. Alam \etal}.
\newblock {\em Phys. Rev. Lett.}, 74:2885, 1995.

\bibitem{DELPHIMH}
{DELPHI Collab., P. Abreu, \etal}.
\newblock {\em {CERN-PPE/{\bf{97-145}}}}, 1997.
\newblock submitted to Phys. Lett. {\bf{B}}.

\bibitem{ABE97A}
{CDF Collab., F. Abe \etal}.
\newblock {\em Phys. Rev. Lett.}, 79:357, 1997.

\bibitem{CIUCHINI97A}
{M. Ciuchini, G. Degrassi, P. Gambino and G.F. Giudice}.
\newblock {{\bf{CERN-TH}}-97-279, E-print {\bf{hep-ph/9710335}} (unpublished)}.

\end{thebibliography}

\end{document}